\documentclass{svproc}
\usepackage{url}
\usepackage{graphicx}
\usepackage{multicol}
\usepackage{xspace}
\usepackage{amsmath}
\usepackage{hyperref}
\usepackage{comment}
\usepackage[ruled, vlined, linesnumbered]{algorithm2e}
\usepackage{footmisc}

\hypersetup{
    bookmarks=true,         
    unicode=false,          
    pdftoolbar=true,        
    pdfmenubar=true,        
    pdffitwindow=false,     
    pdfnewwindow=true,      
    colorlinks=true,       
    linkcolor=black,          
    citecolor=black,        
    filecolor=black,      
    urlcolor=black           
}

\newcommand{\spara}[1]{\smallskip\noindent{\bf #1}}
\newcommand{\algo}{\textsf{Structural-balance-viz}\xspace}
\newcommand{\laplacian}{\ensuremath{\bar{L}}\xspace}

\DeclareMathOperator*{\diag}{diag}

\newcommand{\revision}[1]{\textcolor{black}{#1}}

\begin{document}
\mainmatter

\title{Visualizing structural balance in signed networks}

\author{Edoardo Galimberti\inst{1,2} \and
Chiara Madeddu\inst{1} \and
\\
Francesco Bonchi\inst{2} \and
Giancarlo Ruffo\inst{1}}
\tocauthor{Edoardo Galimberti,
Chiara Madeddu,
Francesco Bonchi and
Giancarlo Ruffo}
\authorrunning{Edoardo Galimberti et al.}

\institute{University of Turin, Italy \and
ISI Foundation, Italy}

\maketitle
\sloppy

\begin{abstract}
\emph{Network visualization} has established as a key complement to network analysis since the large variety of existing network layouts are able to graphically highlight different properties of networks.
However, \emph{signed networks}, i.e., networks whose edges are labeled as friendly (positive) or antagonistic (negative), are target of few of such layouts and none, to our knowledge, is able to show \emph{structural balance}, i.e., the tendency of cycles towards including an even number of negative edges, which is a well-known theory for studying friction and polarization.

In this work we present \algo: a novel visualization method showing whether a \revision{connected signed network} is balanced or not and, in the latter case, how close the network is to be balanced.
\algo exploits spectral computations of the signed Laplacian matrix to place network's nodes in a Cartesian coordinate system resembling a balance (a scale). Moreover, it uses edge coloring and bundling to distinguish positive and negative interactions.
The proposed visualization method has characteristics desirable in a variety of network analysis tasks:
\algo is able to provide indications of balance/polarization of the whole network and of each node,
to identify two factions of nodes on the basis of their polarization,
and to show their cumulative characteristics.
Moreover, the layout is reproducible and easy to compare.
\algo is validated over synthetic-generated networks and applied to a real-world dataset about political debates confirming that it is able to provide meaningful interpretations.

\end{abstract}

\section{Introduction}
\label{sec:intro}

\begin{figure}[t]
\vspace{-4mm}
\centerline{
\begin{tabular}{cc}
\includegraphics[width=0.3\columnwidth]{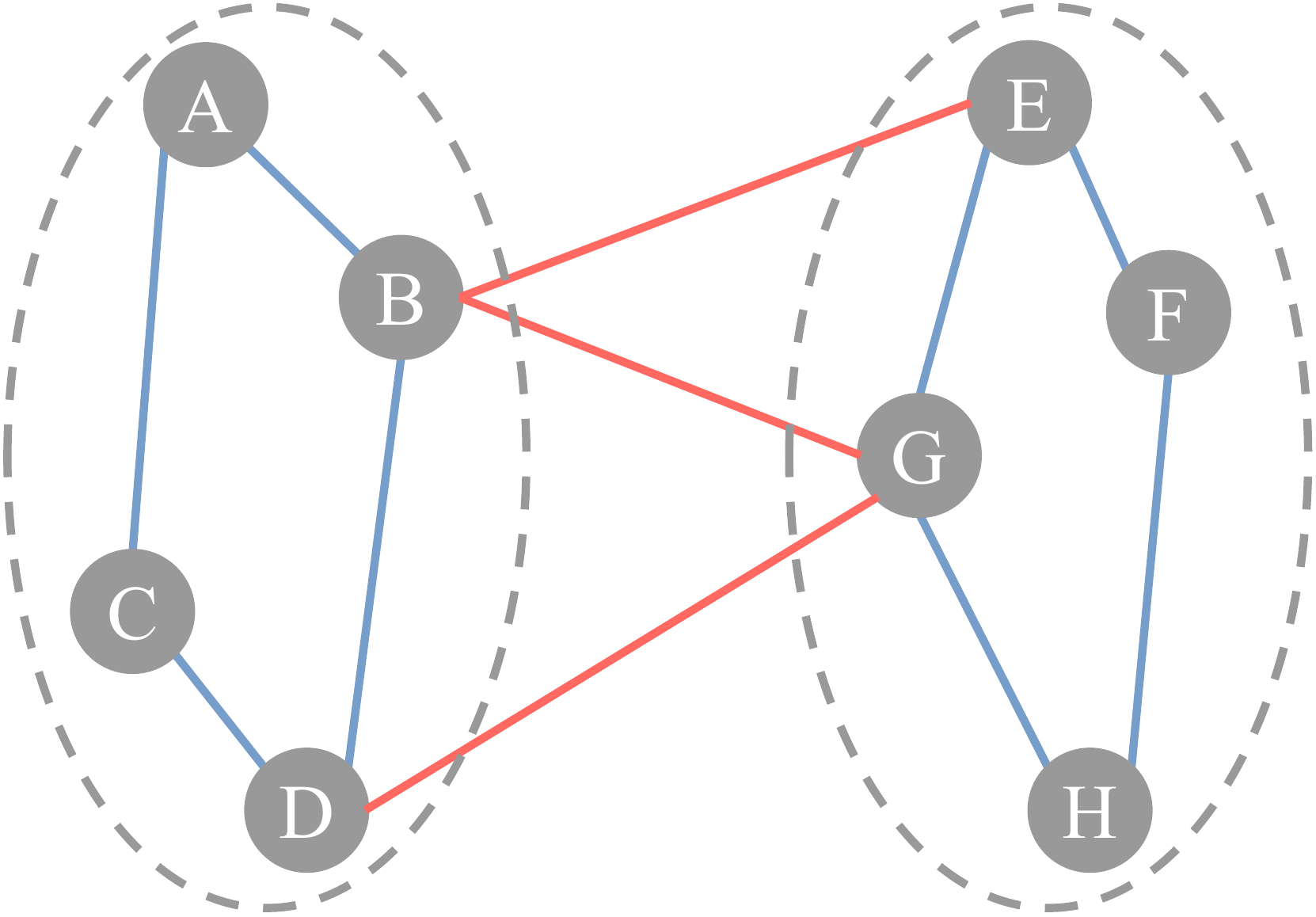} \hspace{0.6cm} & \hspace{0.6cm}\includegraphics[width=0.3\columnwidth]{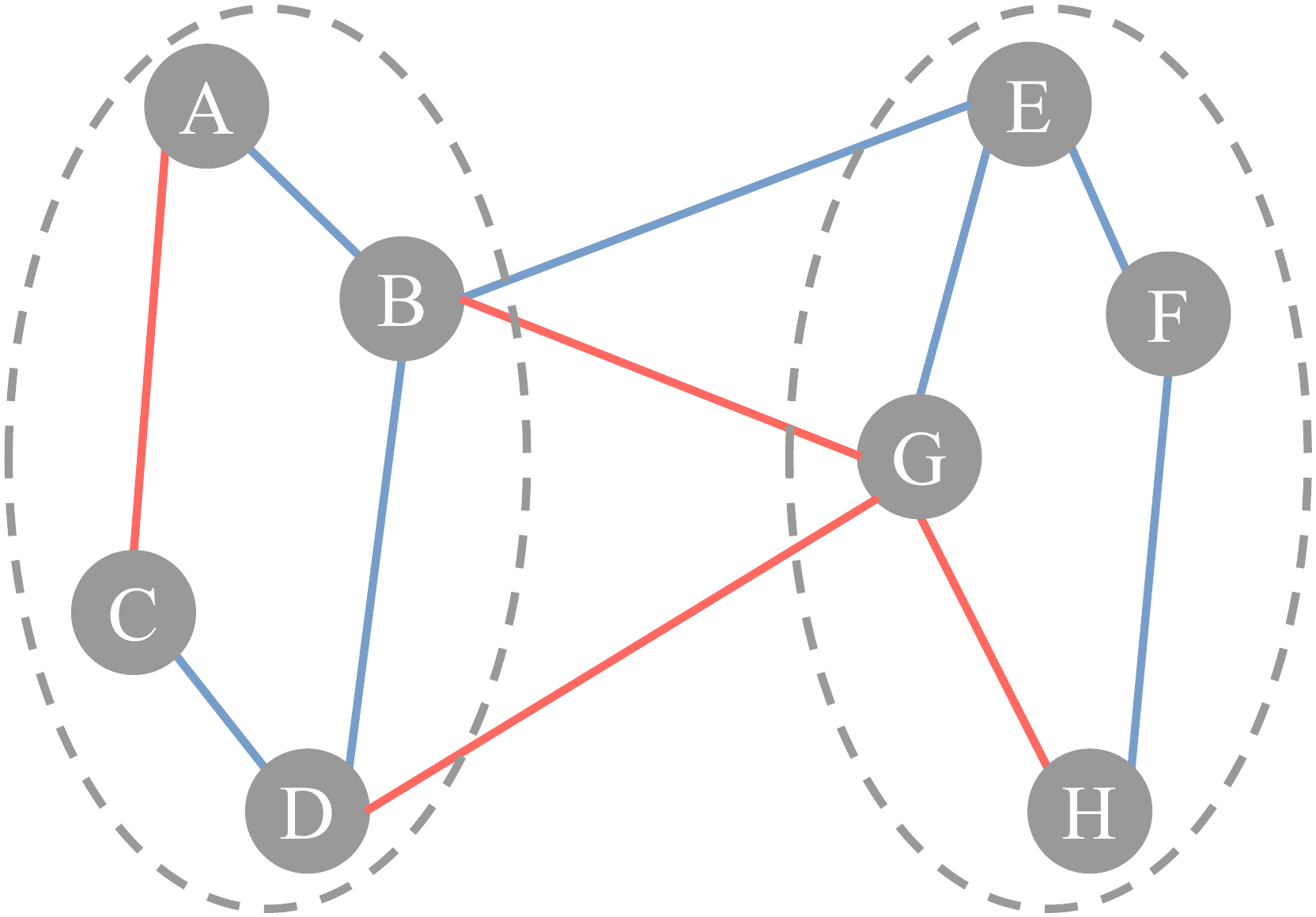} \\
balanced & unbalanced \\
\end{tabular}
}
\caption{Examples of balanced (left) and unbalanced (right) networks.
Positive edges are reported in blue, while negative edges in red.}
\label{fig:examples_intro}
\vspace{-4mm}
\end{figure}

\emph{Signed networks} are simple yet informative network representations in which edges are annotated as positive or negative~\cite{harary1953notion}.
They are applied in a large variety of domains in which interactions between entities are either friendly or antagonistic, e.g.,
international relations~\cite{doreian2015structural}, 
and online social media and social networks~\cite{tang2016survey}.
The theory of \emph{structural balance} has established as the standard for studying, from a theoretical standpoint in sociology and psychology, the formation of opinions in both individuals and social groups. 
Structural balance is widely applied to signed networks, e.g., 
for the analysis of social media~\cite{leskovec2010signed},
and the study of opinion separation~\cite{xia2015structural}.
A signed network has been proved to be \emph{structurally balanced} or \emph{balanced} if and only if all cycles are balanced, i.e., include an even number of negative edges~\cite{cartwright1956structural}.
As a consequence, network's nodes can be assigned to two different sets such that we find only positive ties between nodes in the same set and all negative ones between nodes of different sets~\cite{easley2010positive}.
Figure~\ref{fig:examples_intro} shows two simple examples of balanced and unbalanced networks.
The network on the left is balanced and has the two properties discussed above, i.e., all cycles are balanced and a clustering can be found in agreement to all edges' signs.
On the other hand, the network on the right is not balanced: there are unbalanced cycles (e.g., the one composed by the node sequence $[A, B, D, C, A]$) and there are edges disagreeing with the clustering (e.g., edge $(B,E)$).
Even if a balanced network represents the most natural configuration, structural balance is not necessarily a ``positive'' configuration,
e.g., it is observed in the alliance network between European nations just before World War I~\cite{redner2006social}.
Moreover, most of the large real-world networks are expected to be unbalanced since a single unbalanced cycle makes the whole network unbalanced.
Therefore, it has also been shown the importance of measuring to what extent an unbalanced signed network is close to be balanced~\cite{kunegis2010spectral}.
Structural balance is also linked with \emph{group polarization}, i.e., the division of a group of entities (e.g., nodes of a network) into two subgroups each reaching consensus and having opposite opinions~\cite{bonchi2019discovering}.

\emph{Network visualization} has emerged as a key complement to standard network analysis techniques to fill the gap between computation and interpretation, communicate findings, and deepen insight~\cite{krzywinski2011hive}.
A large variety of network layouts exists in literature~\cite{kaufmann2003drawing}
and, also, implemented for visualization applications, as, e.g., Gephi and Cytoscape. 
Surprisingly, little attention has been paid to the visualization of signed networks~\cite{kunegis2010spectral}
and, to our knowledge, none of the existing layouts highlights structural balance properties of signed networks.

In this work we tackle the task of identifying, through a visualization, whether a \revision{connected signed network} is balanced or unbalanced and, in the latter case, how much the network is unbalanced.
The proposed visualization method, \algo, places nodes in a Cartesian coordinate system exploiting spectral properties of the signed Laplacian matrix.
Edges are colored and bundled to make positive and negative signs distinguishable and to ease the understanding of the global balance/polarization of the network.
At a glance, it is possible to catch if a network is balanced: no positive edges cross the $y$-axis and no negative edges have both endpoints in the same quadrant, namely, the $y$-axis finds a partition of the nodes as explained in~\cite{easley2010positive}.
The visual perception of the portion of edges ``disagreeing'' with the partitioning, i.e., the fraction of positive edges crossing the $y$-axis and negative edges internal to a quadrant, gives an indication of the level of balance of a network.
Moreover, we utilize the $x$-axis as a scale to show cumulative characteristics of the sets of nodes identified by the $y$-axis,
and include a textual indication of the level of balance of the network under analysis in order to improve the comparability between different visualizations.

The layout produced by \algo has the following characteristics that are useful in a variety of network analysis tasks:
$(i)$ it shows whether the input network is balanced or not and, in the second case, how close the network is to be balanced;
$(ii)$ by nodes' $x$-coordinate, it provides an indication of the contribute to the balance structure of the network and, also, of the individual balance/polarization of each node (such information might be exploited, e.g., for the task of finding non-polarized representatives~\cite{ordozgoiti2019reconciliation});
$(iii)$ it identifies two factions of nodes on the basis of their polarization which finds applications in clustering problems, e.g., 2-correlation-clustering~\cite{coleman2008local,aref2018computing};
$(iv)$ the scale represented by the $x$-axis shows cumulative characteristics of the identified factions, e.g., size or internal clustering coefficient;
and, $(v)$ the resulting visualization are reproducible (desirable feature but not common to all network layouts, e.g., force based) and easy to compare in terms of balance structure.
We verify such characteristics by running \algo on synthetic networks and a real-world dataset representing political debates.

The rest of the paper is structured as follows.
Section~\ref{sec:related} covers the related work, in Section~\ref{sec:algorithm} we describe our visualization procedure, while Section~\ref{sec:applications} shows the experimental validation and the real-world application.
Finally, Section~\ref{sec:conclusions} concludes the paper.

\section{Related work}
\label{sec:related}

\spara{Structural balance and signed networks.}
The concept of \emph{structural balance} first appears as psychological theory of balance in triangles of sentiments.
\emph{Signed networks} are later introduced in the seminal work by Harary~\cite{harary1953notion}, who also generalizes the balance theory to signed networks~\cite{cartwright1956structural}.
Harary and Kabell develop a simple algorithm to test whether a given signed network is balanced~\cite{harary1980simple} by enumerating the cycles in the network containing an even number of negative edges.
\revision{A complete signed network is balanced if and only if all its triangles are balanced~\cite{easley2010positive}.}
Akiyama et al.~\cite{akiyama1981balancing} study how to estimate the minimum number of sign changes required so that a signed network satisfies the balance property.
Recent works link spectral properties of signed networks to the balance theory.
Hou et al.~\cite{hou2003laplacian} prove that a signed network is balanced if and only if the smallest eigenvalue of the signed Laplacian is~0.
Moreover, \cite{hou2005bounds} investigates the relationship between the smallest eigenvalue of the signed Laplacian and the level of balance of a signed network.

A fundamental problem studied in signed networks is correlation clustering~\cite{bansal2004correlation}, i.e., partition the nodes into clusters so as to maximize (minimize) the number of edges that ``agree'' (``disagree'') with the partitioning.
\revision{The 2-correlation-clustering problem~\cite{coleman2008local}, also known as the frustration-index problem~\cite{aref2018computing}, is also widely studied.}
Finally, a more recent line of work introduces the problem of discovering antagonistic communities in signed networks~\cite{bonchi2019discovering}. 

\spara{Network visualization.}
Many \emph{network visualizations} have been proposed in literature in order to graphically express specific characteristics, properties, and patterns of networks. 
Force-based visualizations map an energy function to the desired layout and minimize it~\cite{kaufmann2003drawing}.
Hive plots~\cite{krzywinski2011hive} place nodes on radially oriented linear axes according to a coordinate system defined by nodes characteristics and/or network properties.
Eigenvectors are exploited for visualizing networks in different works.
In particular,~\cite{kunegis2010spectral} studies the application of clustering, prediction, and visualization methods to signed networks by using the signed Laplacian and its eigenvalue decomposition.
Despite using eigenvectors to place nodes in a Cartesian coordinate system, the visualization algorithm of~\cite{kunegis2010spectral} has different purposes and strongly differs from ours:
$(i)$ it wants to highlight clustering properties and not structural balance;
$(ii)$ it does not provide information about the contribute of each node to the balance/polarization structure;
$(iii)$ it does not cluster nodes into two factions and, therefore, it cannot show factions' cumulative properties;
and, $(iv)$ the resulting layouts are hardly comparable between each other.

\section{Visualizing structural balance}
\label{sec:algorithm}

In this section we describe the details of \algo, the proposed visualization method whose main objective is to show whether a \revision{connected signed network} is balanced or unbalanced and, in the latter case, how much the network is unbalanced.

First, we provide preliminary notations and definitions.
We denote a signed undirected network as $G = (V, E_+, E_-)$, where $V$ is a set of nodes, $E_+$ is a set of positive edges, and $E_-$ is a set of negative edges.
\revision{In this work, we require $G$ to be connected.}
Let $A$ be the signed adjacency matrix of $G$, i.e., for each pair of nodes $u,v \in V$, $A[u,v] = 1$ if $(u,v) \in E_+$, $A[u,v] = -1$ if $(u,v) \in E_-$, $A[u,v] = 0$ otherwise.
Let also $\bar{D} = \diag(\bar{d}_{u_1},\ldots,\bar{d}_{u_{|V|}})$ be the signed degree matrix of $G$, where $\bar{d}_{u} = \sum_{v \in V} |A[u,v]|$ represents the signed degree, i.e., the number of neighbors disregarding the sign, of a node $u \in V$.
Finally, we define the signed Laplacian matrix of $G$ as:
$$
\laplacian = \bar{D} - A.
$$

\begin{algorithm}[t]
\DontPrintSemicolon
\newcommand\commentfont[1]{\small\ttfamily{#1}}
\SetCommentSty{commentfont}

\KwIn{A signed network $G = (V, E_+, E_-)$ and a network measure $\mu$ (optional).}
\KwOut{A visualization of $G$.}

\tcc{Eigenvalue decomposition}
compute the signed Laplacian \laplacian of $G$\;
compute the smallest eigenvalue $\lambda_m$ of \laplacian and its corresponding eigenvector $\vec{v}_m$\; \label{line:algo:eigen}

\tcc{Nodes coordinates}
$\mathbf{X} \leftarrow \emptyset$; \ \ $\mathbf{Y} \leftarrow \emptyset$\;
\ForAll{$u \in V$}
{
	\label{line:algo:for}
	$\mathbf{X}[u] = \vec{v}_m[u]$\;
	$\mathbf{Y}[u] = |\{v \in V \mid \vec{v}_m[v] = \vec{v}_m[u] \land v < u\}|$\;
}

\tcc{Edge partitioning}
$E_+^i = \{e = (u,v) \in E_+ \mid \mathbf{X}[u] = \mathbf{X}[v]\}$\; \label{line:algo:edgestart}
$E_-^i = \{e = (u,v) \in E_- \mid \mathbf{X}[u] = \mathbf{X}[v]\}$\; 
$E_+^e = E_+ \setminus E_+^i$\;
$E_-^e = E_- \setminus E_-^i$\; \label{line:algo:edgeend}

\tcc{Drawing}
draw the Cartesian axes\; \label{line:algo:draw}
draw the nodes in $V$ according to $\mathbf{X}$ and $\mathbf{Y}$\;
draw the edges in $E_+^i$ in \textbf{blue} with \textbf{horizontal-external} bundling\;
draw the edges in $E_-^i$ in \textbf{red} with \textbf{horizontal-internal} bundling\;
draw the edges in $E_+^e$ in \textbf{blue} with \textbf{vertical-upper} bundling\;
draw the edges in $E_-^e$ in \textbf{red} with \textbf{vertical-lower} bundling\; \label{line:algo:drawend}

\tcc{Additional features}
\If{$\mu \neq \textsc{null}$}
{
	\label{line:algo:feature1}
	$C_l = \{u \in V \mid \mathbf{X}[u] < 0\}$; \ \ $C_r = \{u \in V \mid \mathbf{X}[u] \geq 0\}$\; \label{line:algo:comm}
	let $\gamma = \mu(C_l) - \mu(C_r)$ be the angular coefficient of the $x$-axis\;
}
draw the label ``$y = \lambda_m$''\; \label{line:algo:feature2}

\caption{\algo}\label{alg:algo}
\end{algorithm}

We now describe our algorithm for visualizing structural balance in signed networks, which is outlined as Algorithm~\ref{alg:algo}.
As mentioned beforehand, \algo makes use of the signed Laplacian of the input network $G$.
In fact, it starts by computing the signed Laplacian together with its smallest eigenvalue $\lambda_m$ and the corresponding eigenvector $\vec{v}_m$ (Line~\ref{line:algo:eigen}).
At this point, we already have all the information required for the visualization handy. 
At first, we identify the coordinates of the nodes in $V$ and store them in $\mathbf{X}$ and $\mathbf{Y}$ (cycle starting at Line~\ref{line:algo:for}).
The $x$-coordinate of each node $u$ is directly obtained by the element of $\vec{v}_m$ corresponding to $u$.
Since more than a node might have the same abscissa and we want to avoid nodes to overlap, the $y$-coordinates are computed in order to distribute nodes having the same $x$-coordinate vertically.
Next (Lines~\ref{line:algo:edgestart}~-~\ref{line:algo:edgeend}), edges are divided into four sets since, on the basis of the coordinates of their endpoints and of their sign, different layouts are applied:
\begin{itemize}
\item $E_+^i$ contains the positive edges having two endpoint with the same $x$-coordinate;
\item $E_-^i$ contains the negative edges having two endpoint with the same $x$-coordinate;
\item $E_+^e$ contains the positive edges having two endpoint with different $x$-coordinate;
\item $E_-^e$ contains the negative edges having two endpoint with different $x$-coordinate.
\end{itemize}
\algo is then ready to draw the visualization (Lines~\ref{line:algo:draw}~-~\ref{line:algo:drawend}).
At first, the Cartesian axes and the nodes are positioned.
Then, the edges are drawn exploiting coloring and bundling to highlight their sign.
In particular, positive edges are depicted in blue, while negative edges in red.
A positive edge $e_+ \in E_+$ is bundled towards the top of the visualization, if $e_+ \in E_+^e$, or externally, if $e_+ \in E_+^i$;
while a negative edge $e_- \in E_-$ is bundled towards the bottom, if $e_- \in E_-^e$, or internally, if $e_- \in E_-^i$.

In order to improve the informativeness of our layout, we include two additional features in \algo (from Line~\ref{line:algo:feature1}):
one wants to provide information about the two sets of nodes identified by the $y$-axis,
while the latter has the aim of making different visualizations more comparable.

Any eigenvector $\vec{v}$ of the signed Laplacian can be used to derive a partition of network's nodes into two sets on the basis of the sign of the corresponding elements in $\vec{v}$.
\revision{Such partitioning is at the basis of spectral-clustering methods~\cite{coleman2008spectral} and it can identify polarized structures, i.e., two sets of nodes showing high internal consensus and warring between each other~\cite{bonchi2019discovering}.} 
In the proposed visualization, the two sets are identified by the nodes in the left and in the right quadrants, i.e., $C_l$ and $C_r$ computed at Line~\ref{line:algo:comm} of \algo, respectively.
In practical applications, it is often of interest to know (and visualize) network measures of the two polarized sets, e.g., size, internal clustering coefficient, internal density of positive edges, ratio of positive edges, etc.
We provide a simple visual expedient based on the angular coefficient of the $x$-axis that resembles the behavior of a scale.
Let $\mu$ be the network measure of interest.
Note that $\mu$ is an optional input parameter of \algo and the lines corresponding to this additional feature are executed if $\mu$ is actually provided in input.
We define the angular coefficient of the $x$-axis as
$$
\gamma = \mu(C_l) - \mu(C_r).
$$

The work enclosed in~\cite{hou2003laplacian,hou2005bounds} proves theoretical bounds on the smallest eigenvalue of the Laplacian of a signed network and investigates its relationship with respect to the level of balance in the network.
It is shown that a \revision{connected signed network} is structurally balanced if and only if $\lambda_m = 0$, i.e., the smallest eigenvalue of the Laplacian is zero, and that the higher $\lambda_m$, the lower the level of balance of the network is.
\revision{Therefore, $\lambda_m$ is the simplest indicator to take into account for comparing structural balance in different networks (of equal densities).
More complex indicators of balance could also be employed~\cite{aref2017measuring}.
}
Ideally, the $y$-coordinate where the $x$-axis crosses the $y$-axis would be a simple manner to graphically show $\lambda_m$.
Unfortunately, we devoted consistent effort to visualize such information in this way, but all attempts worsened the clarity of the layout (e.g., cut off edges).
To this extent, we include in \algo a label reporting the value of $\lambda_m$ on the top of the $y$-axis and leave the visualization of $\lambda_m$ without the label as future work.

The time complexity of \algo is governed by the time required by the eigenvalue decomposition of \laplacian, while the space complexity is $\mathcal{O}(|V|^2)$, again imposed by \laplacian.
Note that computational-intensive network measures $\mu$ might considerably extend the running time when drawing large networks.

\begin{figure}[t]
\vspace{-4mm}
\centerline{
\includegraphics[width=0.7\columnwidth]{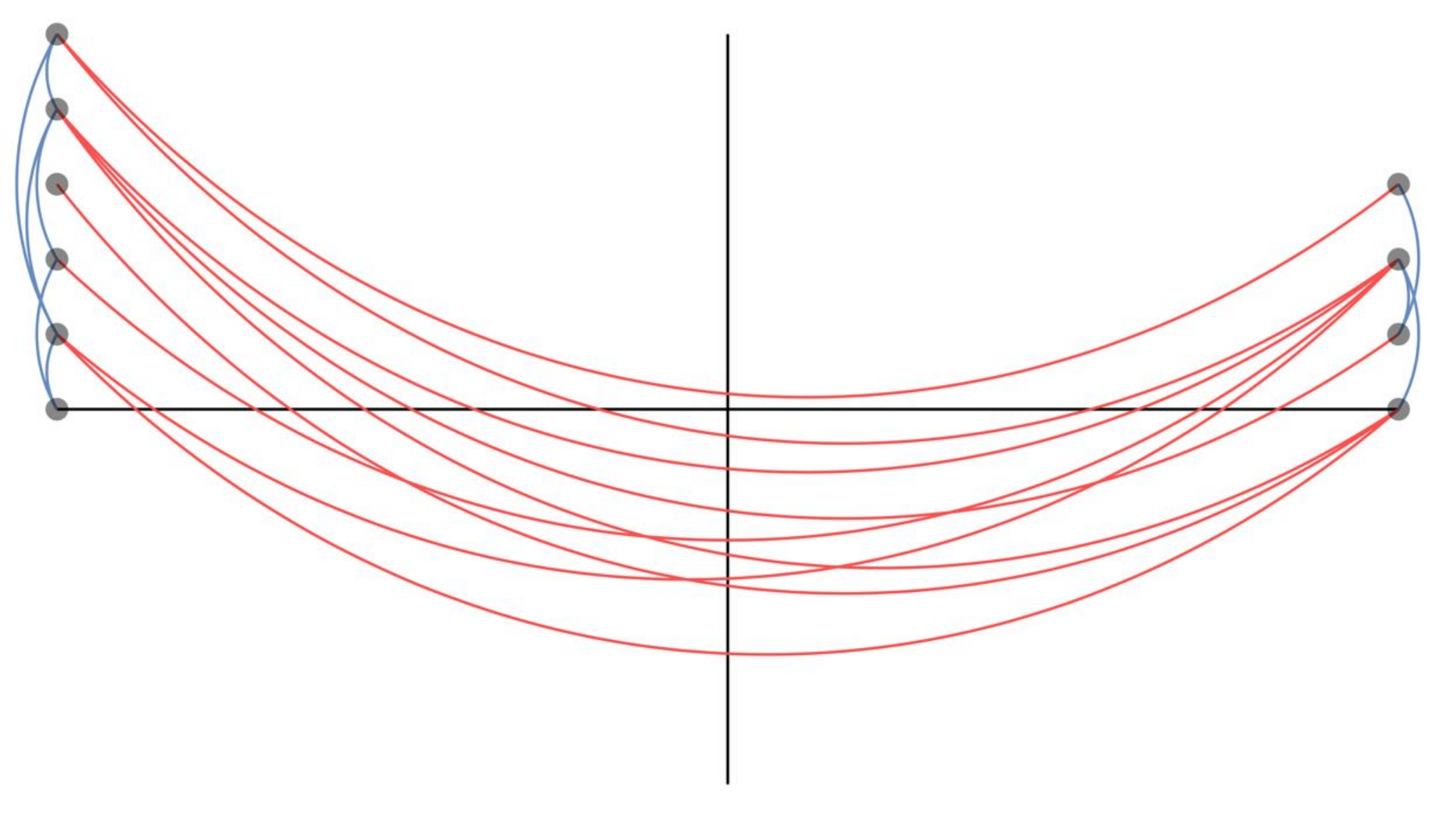}
}
\vspace{-4mm}
\caption{Visualization by \algo of a balanced network: all the cycles are balanced.}
\label{fig:balanced}
\vspace{-2mm}
\end{figure}

\begin{figure}[t!]
\centerline{
\includegraphics[width=0.7\columnwidth]{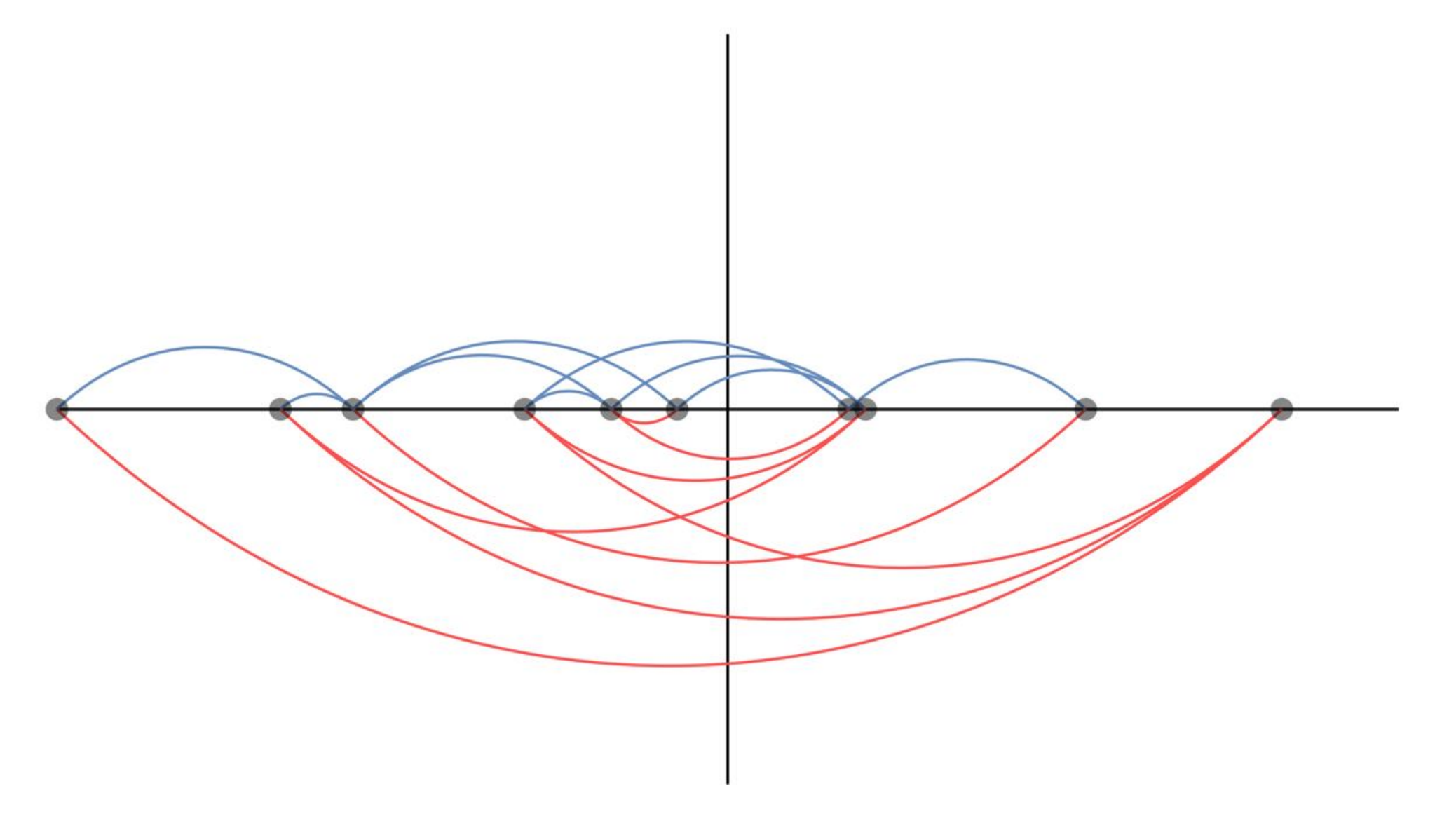}
}
\vspace{-4mm}
\caption{Visualization by \algo of an unbalanced network: not all the cycles are balanced.}
\label{fig:unbalanced}
\vspace{-4mm}
\end{figure}

Figures~\ref{fig:balanced}~and~\ref{fig:unbalanced} show two examples of visualizations generated by \algo for a balanced and an unbalanced network, respectively.
For such visualizations, we remove the label reporting $\lambda_m$ to prove how obvious the difference between the two networks is even without textual information.
\revision{Also, as for all other examples in this paper, edge bundling is not applied;
however, it will be available in the tools we plan to publicly release.}
It is immediate to note that the network represented in Figure~\ref{fig:balanced} is balanced: all the nodes are at the extremes of the $x$-axis and no blue (red) edge crosses the $y$-axis (lays in the same quadrant).
This configuration highlights the fact that all the cycles of the represented network are balanced.
On the other hand, Figure~\ref{fig:unbalanced} shows an unbalanced network since there are positive edges in-between the two factions of nodes and a negative edge within two nodes in the left quadrant; therefore, we easily find the presence of unbalanced cycles.

\begin{figure}[t]
\vspace{-4mm}
\centerline{
\includegraphics[width=0.7\columnwidth]{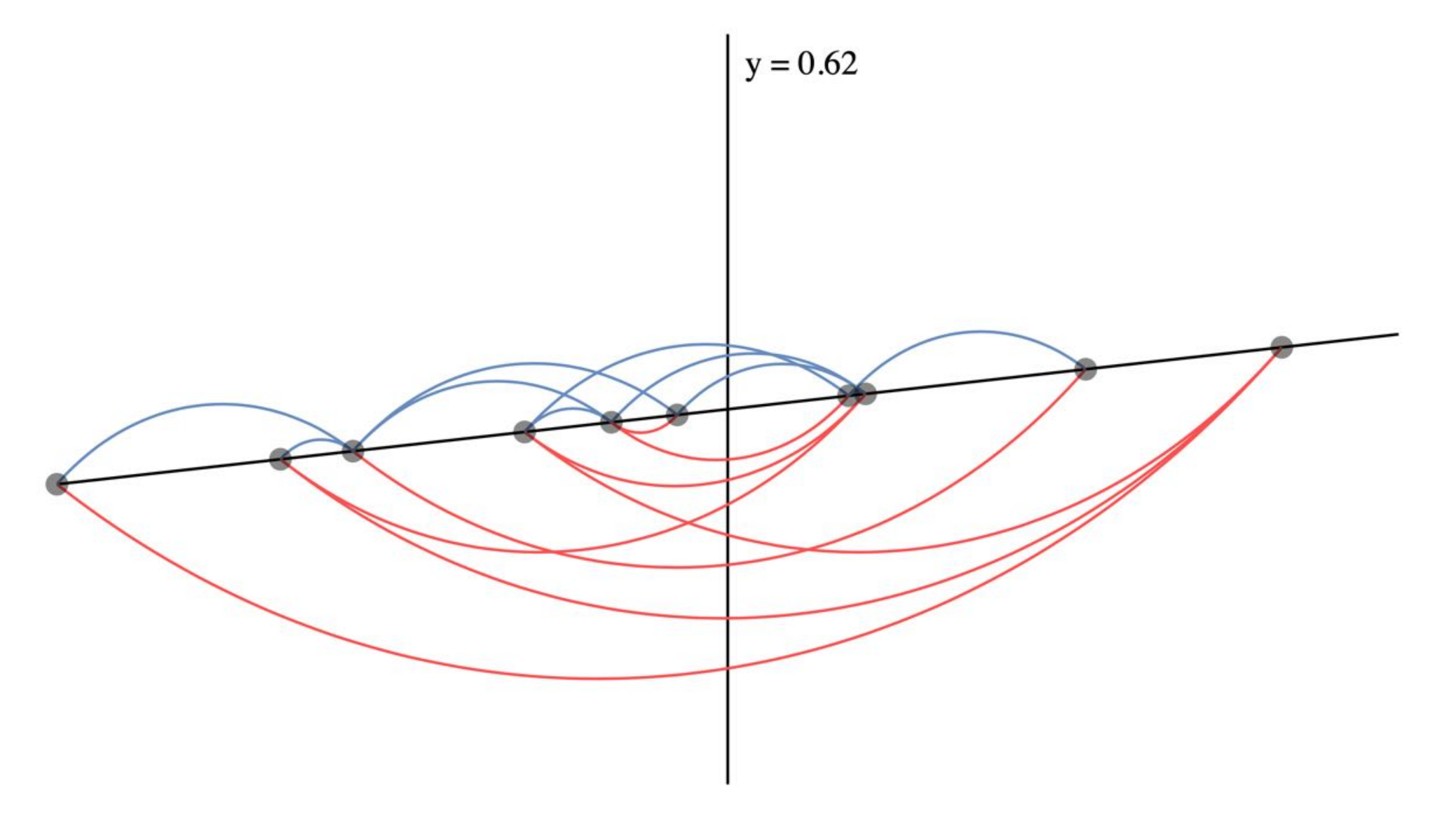}
}
\vspace{-4mm}
\caption{Visualization by \algo of an unbalanced network with the the additional features.
The $x$-axis scale compares the size of the two factions of nodes.}
\label{fig:enhancements}
\vspace{-4mm}
\end{figure}

Figure~\ref{fig:enhancements} shows the same network of Figure~\ref{fig:unbalanced} with both the additional features of \algo;
in this case, the $x$-axis scale compares the size of the two factions of nodes, i.e., $\mu$ counts the number of nodes in the sets.
At a glance, it is possible to understand that the left faction is slightly larger than the right one (six and four nodes, respectively) and that the smallest eigenvalue of the signed Laplacian is not far from zero; this means that the network is not far from being balanced (i.e., there are not many unbalanced cycles).

\section{Validation and application}
\label{sec:applications}

In this section we validate the proposed network layout by visualizing synthetic networks.
Also, we apply \algo to derive concrete insights from a dataset representing political debates.

We develop \algo by using D3.js with a Java back-end.
The visualization is made available by a web interface that allows the selection of the input dataset and of $\mu$ (i.e., the network measure that defines the angular coefficient of the $x$-axis)\footnote{Code available at \href{https://github.com/egalimberti/balance_visualization}{github.com/egalimberti/balance\_visualization}.}.
The current implementation can consider only the size of the sets of nodes as $\mu$, but the code is easily extendable to consider other characteristics.
The time required by our implementation to produce each visualization has always been less than a few seconds.

\begin{figure}[t]
\vspace{-4mm}
\centerline{
\begin{tabular}{cc}
\includegraphics[width=0.5\columnwidth]{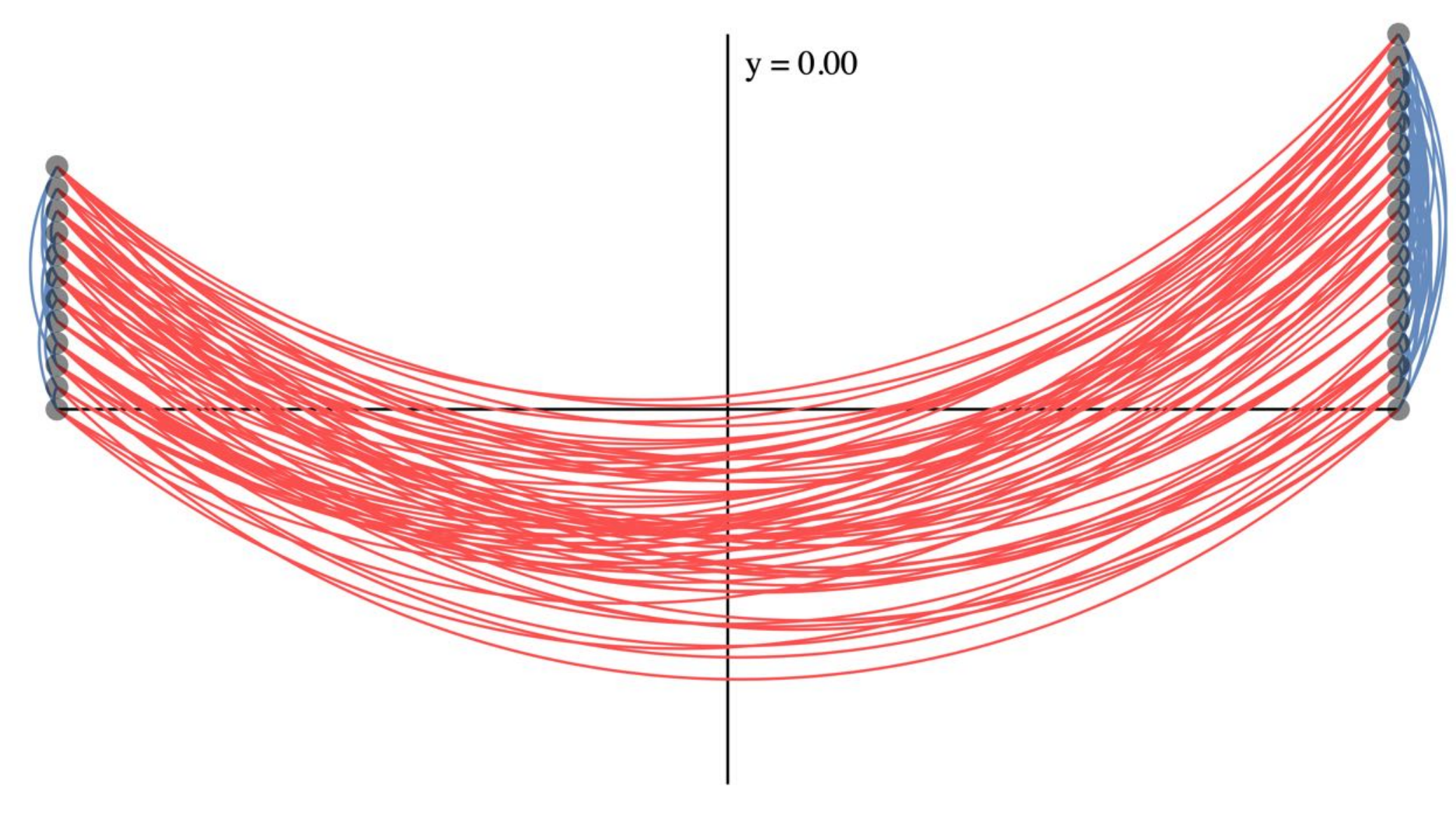} \hspace{0.25cm} & \hspace{0.25cm}\includegraphics[width=0.5\columnwidth]{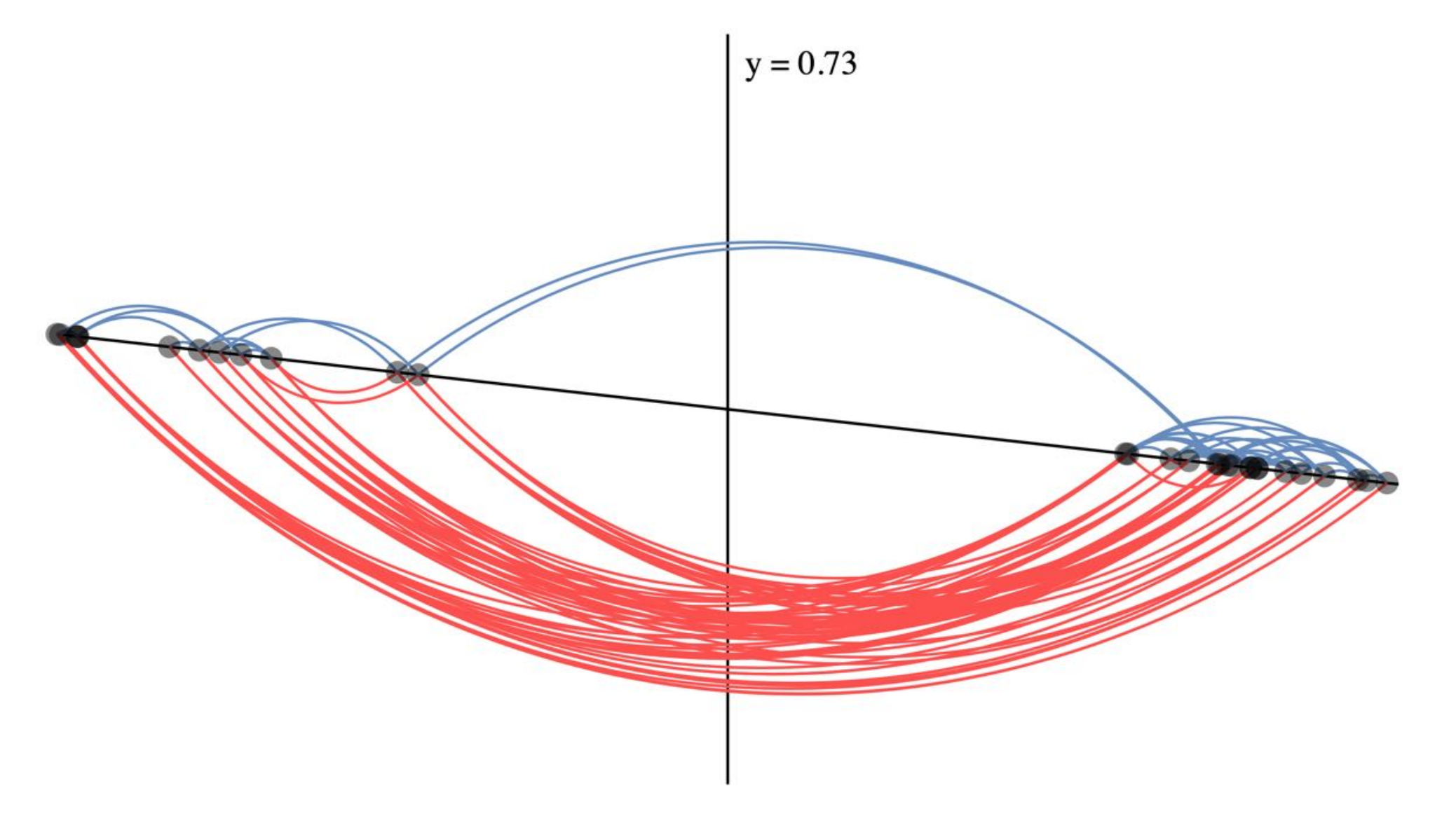} \\
$\nu = 0$ & $\nu = 0.2$ \\
\includegraphics[width=0.5\columnwidth]{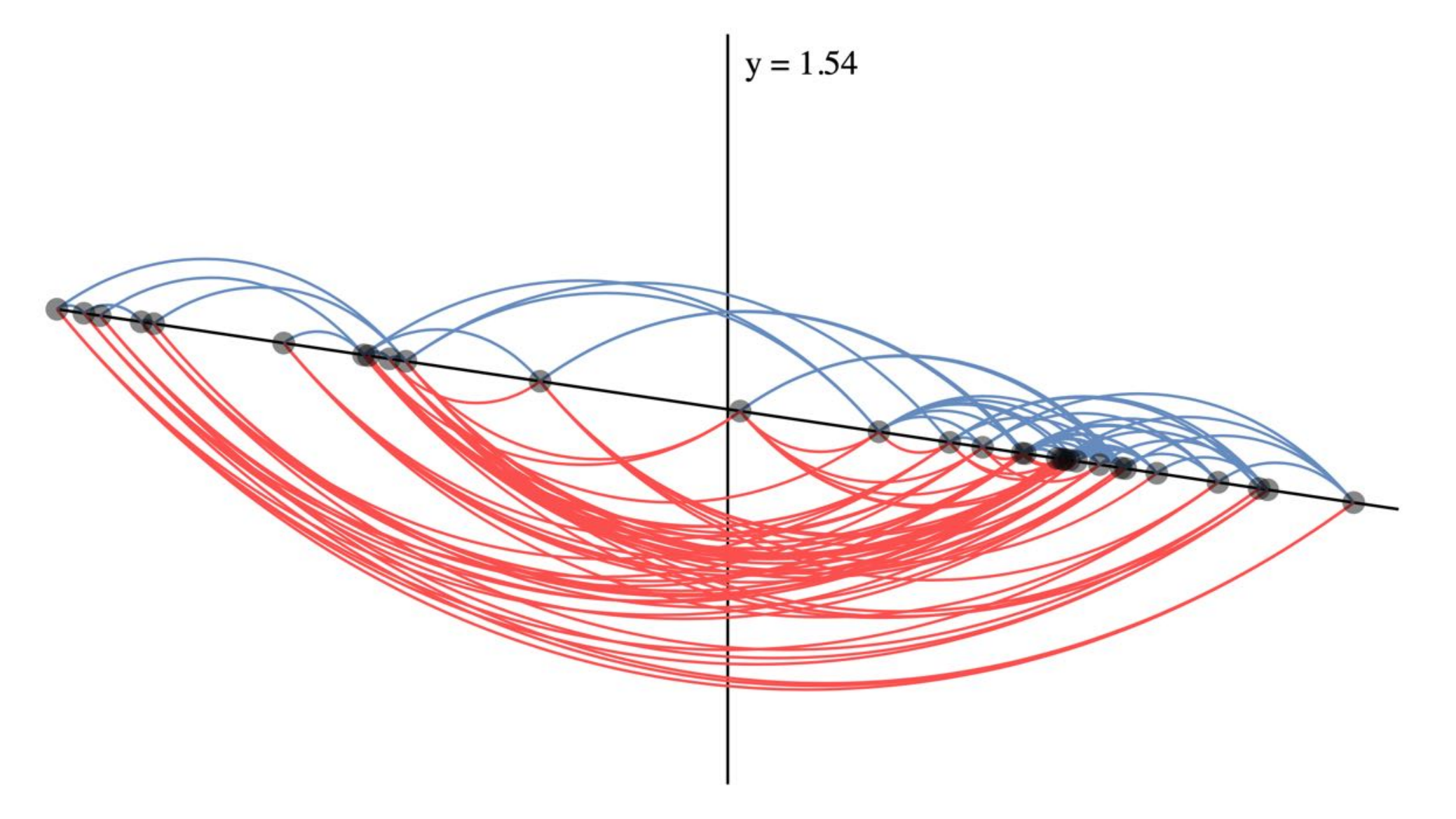} & \includegraphics[width=0.5\columnwidth]{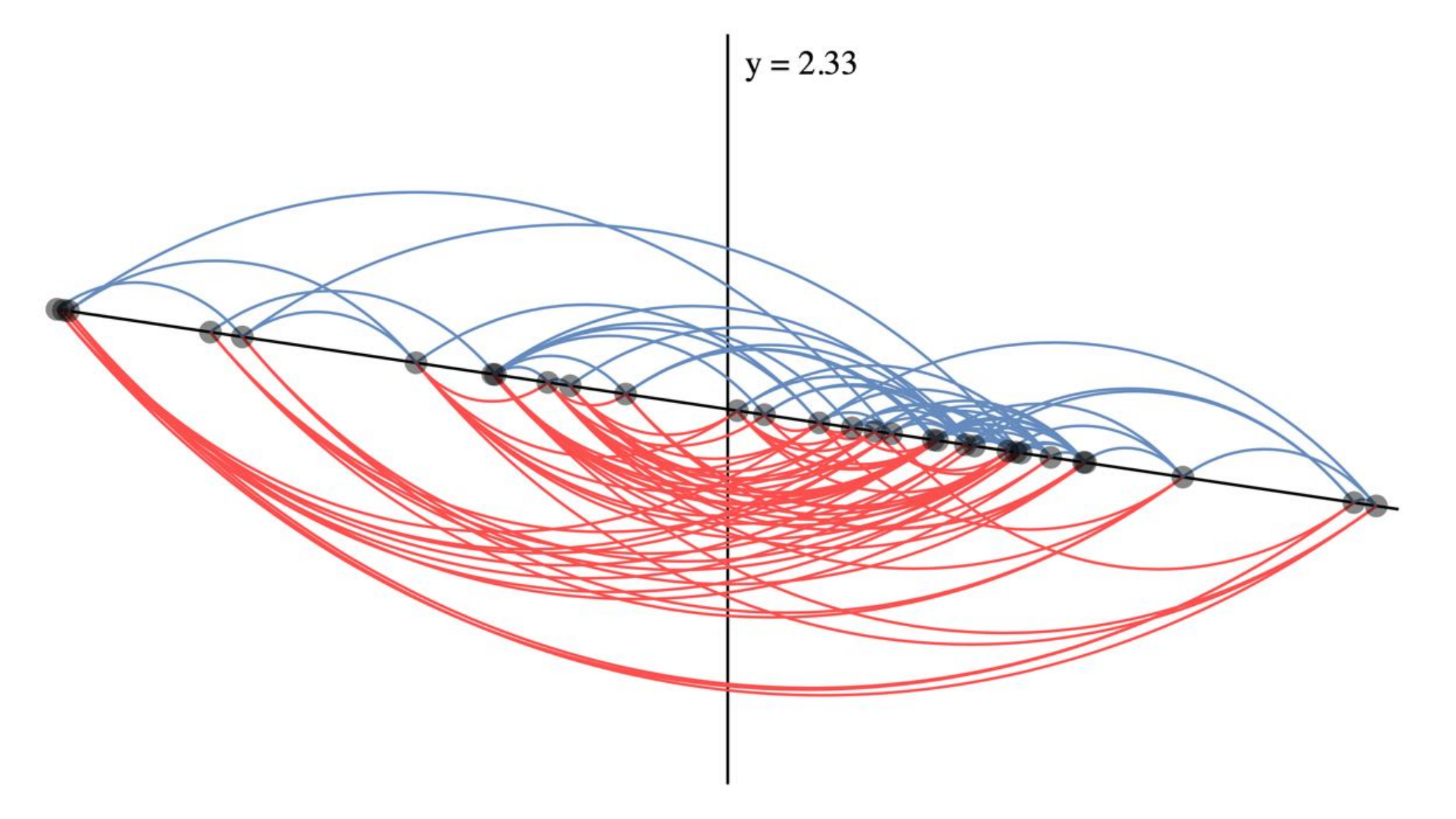} \\
$\nu = 0.4$ & $\nu = 0.6$ \\
\includegraphics[width=0.5\columnwidth]{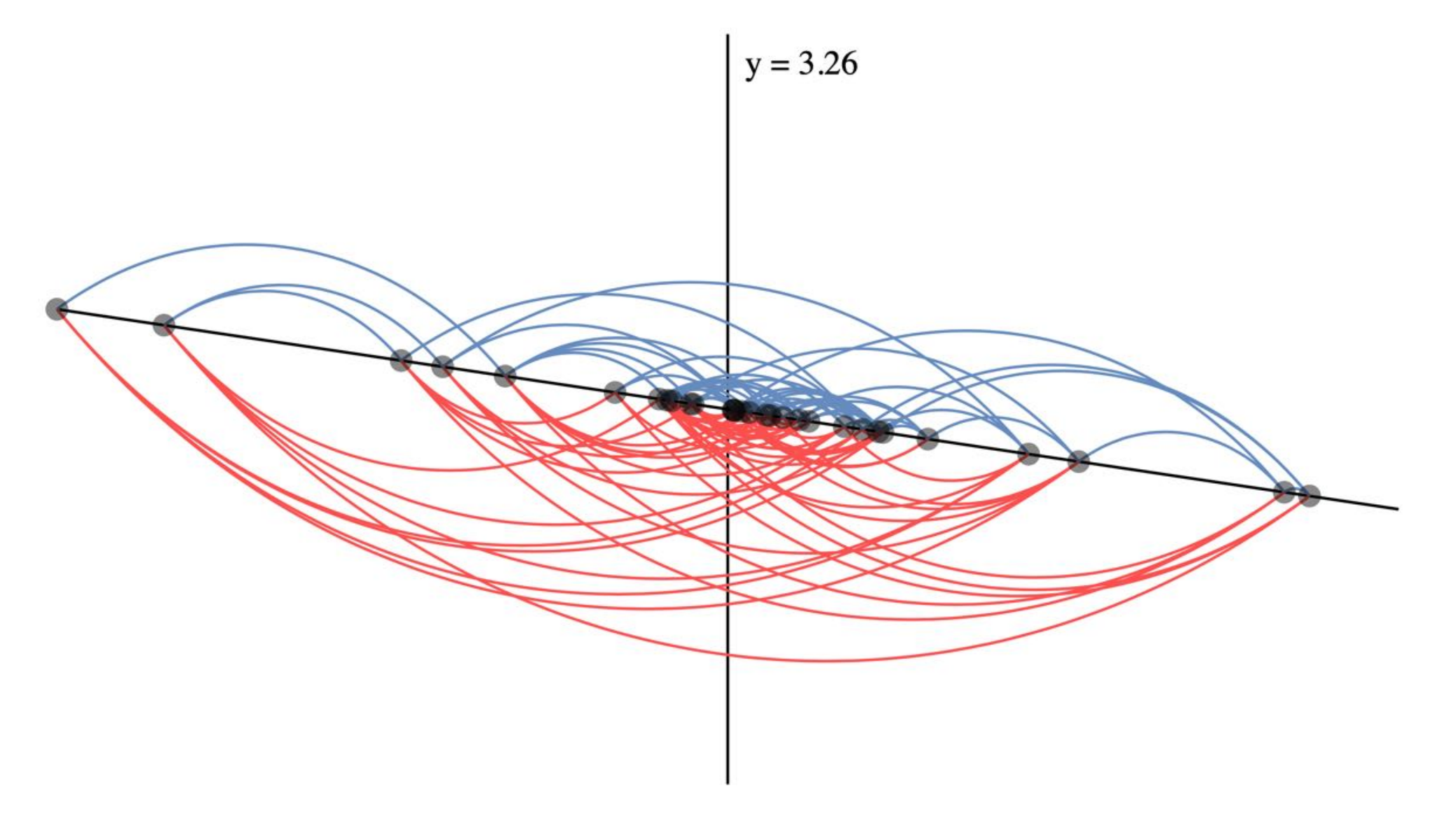} & \includegraphics[width=0.5\columnwidth]{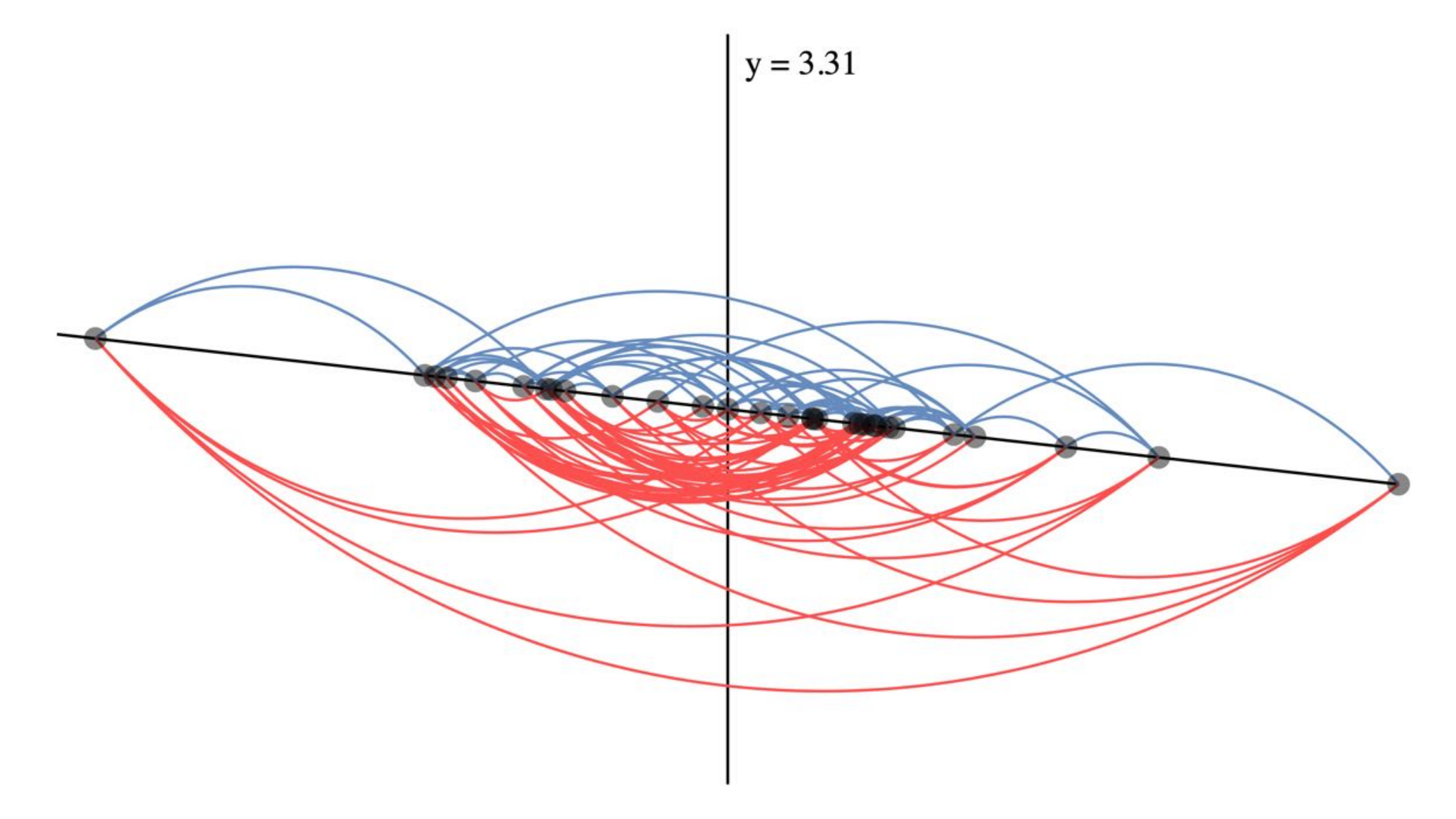} \\
$\nu = 0.8$ & $\nu = 1$ \\
\end{tabular}
}
\caption{Visualization by \algo of synthetic networks for increasing values of $\nu$ ($n = 30$, $\delta = 0.3$).}
\label{fig:exemple}
\vspace{-4mm}
\end{figure}

\subsection*{Validation: synthetic networks}
We first focus our attention on synthetic-generated networks with the aim of proving that the visualizations produced by \algo are easily comparable.
The generative process for signed networks we follow requires in input three parameters: $n$ indicates the number of nodes, $\delta$ defines the edge density, while $\nu$ is the ratio of unbalanced triangles in the network (which is another indicator of how much a network is balanced~\cite{easley2010positive}).
The procedure works as follows:
\begin{itemize}
\item generate a complete balanced network of $n$ nodes
(this can be achieved by partitioning the $n$ nodes into two and by  assigning negative sign to the edges connecting nodes in different sets while positive sign to all others edges);
\item randomly remove edges \revision{that do not disconnect the network} until the edge density is less or equal than $\delta$;
\item randomly change signs of edges appearing in balanced triangles until the ratio of unbalanced triangles is less or equal than $\nu$.
\end{itemize}

In Figure~\ref{fig:exemple} we report our visualization for six networks generated by the described procedure by progressively increasing $\nu$ ($\nu \in [0, 0.2, 0.4, 0.6, 0.8, 1]$) while keeping $n$ and $\delta$ fixed ($n = 30$, $\delta = 0.3$).
Therefore, we have the full range of networks in terms of structural balance: on one extreme ($\nu = 0$) the network is perfectly balanced, on the other ($\nu = 1$) the network has no balanced triangles.
When $\nu = 0$, as expected, we obtain the perfectly distinguishable configuration of balanced networks, where all nodes are in either extremes of the $x$-axis, no positive edge crosses the $y$-axis, and no negative edge entirely lies in the same quadrant.
Note that, for the balanced case, we do not provide in input to \algo any network function $\mu$ since the number of nodes in the sets can be inferred by the height of the two stacks.
As $\nu$ grows, the most of the nodes gradually moves from the extreme ordinates to the center of the plot;
nonetheless, even for $\nu = 1$, we note a few highly-polarized nodes at the margins of the horizontal domain.
In addition, more and more both positive edges cross the $y$-axis and negative edges are within one of the two quadrants.
The additional features result to be extremely useful in these cases.
At first, the scale gives a precise indication that the right faction is larger than the left one for all values of $\nu$.
Also, the smallest eigenvalue of the signed Laplancian, which grows coherently with $\nu$, eases the comparison of visualizations that might appear similar (e.g., $\nu = 0.8$ and $\nu = 1$) and provides a definitive indication about the structural balance of the visualized networks.

\subsection*{A case study: the United States Congress network}
Next, we apply \algo to the analysis of a real-world network obtained from data of the United States Congress modeling a political debate\footnote{Dataset available at \href{http://konect.cc}{konect.cc}.}.
Nodes ($|V| = 219$) are politicians speaking in the Congress, edges ($|E_+ \cup E_-| = 521$) denote that a speaker mentions another speaker, while signs report whether mentions are in support (positive) or opposition (negative).

\begin{figure}[t]
\vspace{-4mm}
\centerline{
\includegraphics[width=0.7\columnwidth]{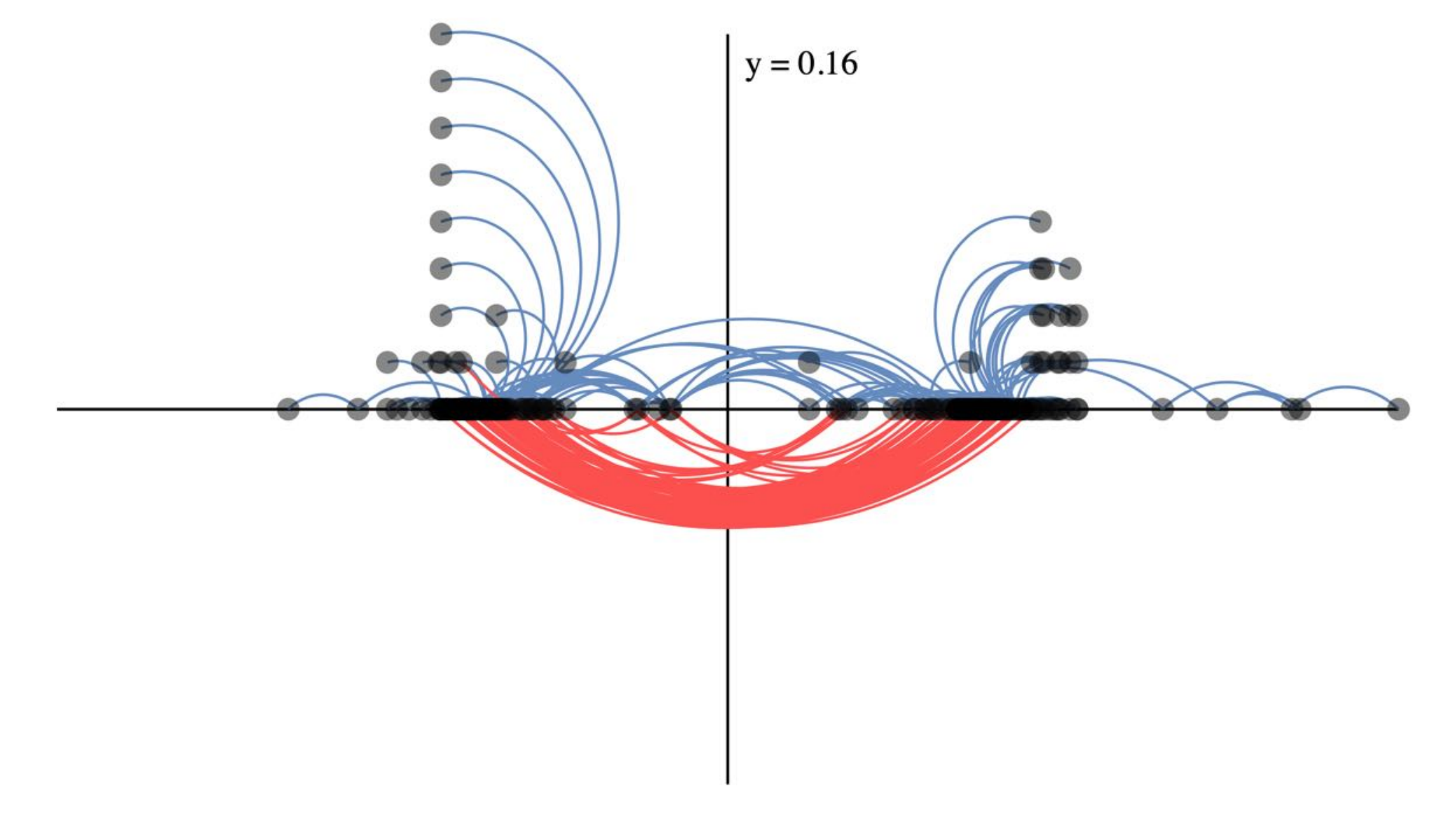}
}
\vspace{-4mm}
\caption{Visualization by \algo of the United States Congress network.}
\label{fig:congress}
\vspace{-4mm}
\end{figure}

\begin{figure}[t]
\centerline{
\includegraphics[width=0.7\columnwidth]{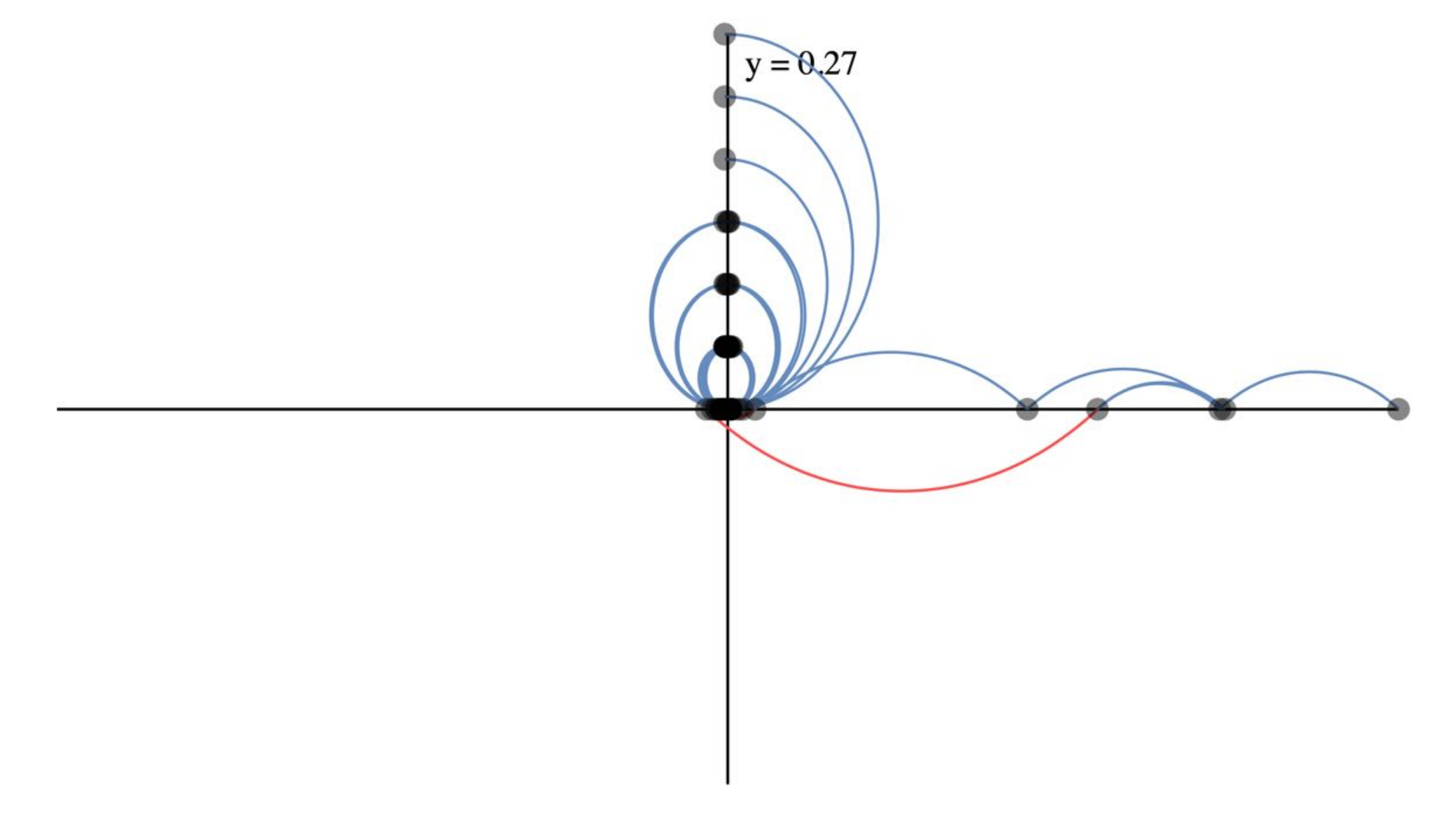}
}
\vspace{-4mm}
\caption{Visualization by \algo of the United States Congress network after sign reshuffling.}
\label{fig:congress_reshuffled}
\vspace{-4mm}
\end{figure}

Figure~\ref{fig:congress} shows the visualization of the original Congress network.
It is easy to notice that the members are divided into two (almost) equally-sized factions that are close to be balanced;
in fact, there is only one negative edge within the left faction and a relatively few positive edges crossing the $y$-axis.
The $x$-axis can be seen as the left-right political spectrum: the most of the politicians are quite moderate, while there are some polarized members especially in the right, and a few nodes close to $x = 0$ (probably the mediators between the two factions).

To have a better understanding of the structural balance of the the Congress network, we compare it to a null model.
\revision{In particular, we maintain the same network structure while reshuffling the edge signs, leaving the number of positive and negative edges unchanged.}
The visualization of the resulting reshuffled network is reported in Figure~\ref{fig:congress_reshuffled}.
In this case, the balance/polarization structure of the network is destroyed since the majority of the nodes collapse close to the origin.
All the negative edges (except one) lay between such nodes and are no more visible in the layout.
Only five members maintain their polarization in the right.
Moreover, the smallest eigenvalue of the signed Laplacian is greater than in the original network.
\revision{All this indications suggests that, the United States Congress network is more balanced/polarized than what is expected by chance, according to a reshuffled null model.}
The Congress is instead quite polarized, very close to being structurally balanced, due to the political parties and alliances.

\section{Conclusions}
\label{sec:conclusions}

In this paper we introduce \algo: a novel algorithm that places nodes in a Cartesian coordinate system, that resembles the behavior of a scale, and exploits edge coloring and bundling for showing whether a \revision{connected signed network} is balance or unbalanced and, in the latter case, how far it is from being balanced.
\algo is validated by the analysis of synthetic networks:
it is proved to provide an indication of balance/polarization of the whole network and individually of each node,
to identify two factions of nodes on the basis of their polarization and show their cumulative characteristics,
and to produce reproducible and easily comparable visualizations.
An application to a real-world dataset about political debates confirms that \algo can provide meaningful insights about the polarization structure of the network.

As future work, we plan to devote more effort in embedding the value of the smallest eigenvalue of the signed Laplacian in \algo without textual supplement.
Moreover, we want to deploy our implementation\revision{, including edge bundling,} to a public web interface and make it available for network visualization tools, e.g., Cytoscape.
Finally, we will employ \algo for future analysis of real-world signed networks.


\end{document}